\documentclass[prl,superscriptaddress,twocolumn]{revtex4}

\usepackage{graphicx}
\usepackage{tabularx}
\usepackage{amsmath}

\newcommand{\rb}[1]{\rm $^{#1}$Rb}
\newcommand{\MHz}{\,{\rm MHz}}
\newcommand{\m}{\,{\rm m}}
\newcommand{\mm}{\,{\rm mm}}
\newcommand{\um}{\,{$\mu$}{\rm m}}
\newcommand{\ms}{\,{\rm ms}}
\newcommand{\ns}{\,{\rm ns}}
\newcommand{\us}{\,\mbox{$\mu$}{\rm s}}
\newcommand{\G}{\,{\rm G}}
\newcommand{\ket}[1]{\mbox{$\left| #1 \right>$}}

\begin{document}

\title{Polarization-controlled single photons}
\author{T. Wilk}
\author{S. C. Webster}
\author{H. P. Specht}
\author{G. Rempe}
\affiliation{Max-Planck-Institut f\"ur Quantenoptik, Hans-Kopfermann-Stra\ss e 1, 85748 Garching, Germany}
\author{A. Kuhn}
\affiliation{Department of Physics, University of Oxford, Clarendon Laboratory, Parks Road, Oxford, OX1 3PU, United Kingdom}

\date{\today}

\begin{abstract}
Vacuum-stimulated Raman transitions are driven between two magnetic substates of a \rb{87} atom strongly coupled to an optical cavity. A magnetic field lifts the degeneracy of these states, and the atom is alternately exposed to laser pulses of two different frequencies. This produces a stream of single photons with alternating circular polarization in a predetermined spatio-temporal mode. MHz repetition rates are possible as no recycling of the atom between photon generations is required. Photon indistinguishability is tested by time-resolved two-photon interference.
\end{abstract}


\maketitle

A major issue in quantum information processing (QIP) is to boost the scale of current experimental implementations so that many quantum bits (qubits) can be handled. One way to get there is to establish a network of stationary quantum systems and to interconnect them by flying qubits such as single photons.  In principle this can be achieved with single atoms coupled to single photons in optical cavities.  Most proposals \cite{97:cirac,99:cabrillo,03:browne,03:duan} require these photons to be indistinguishable, that is in the same spatio-temporal mode of known frequency and polarization.
Previous atom-cavity photon sources \cite{02:kuhn,04:keller,04:mckeever} have met these requirements except for the polarization control, whereas the latter has only been achieved with probabilistic single-photon emitters \cite{86:hong,05:darquie}\footnote{Light emitted by a two-level atom into free space and partially collected by a high N.A. lens is no longer in an angular-momentum eigenstate. This makes it difficult to control the polarization in such a single-photon source.}.  Here we describe the realization of a deterministic single-photon source based on an atom-cavity system which produces photons of known polarization.  A Raman transition between the $m_F=\pm1$ Zeeman sublevels of the $5S_{1/2},F=1$ ground state of a \rb{87} atom is driven by the combination of a pump laser and the cavity vacuum field stimulating the emission of a single photon \cite{06:wilk}.  Such a scheme could also be used to generate a stream of single-photon time-bin qubits in an entangled state.  By combining partial photon production with internal rotations of the atom, photon states such as the GHZ and W states could be constructed \cite{05:schon}.

\begin{figure}
\centering
 \includegraphics{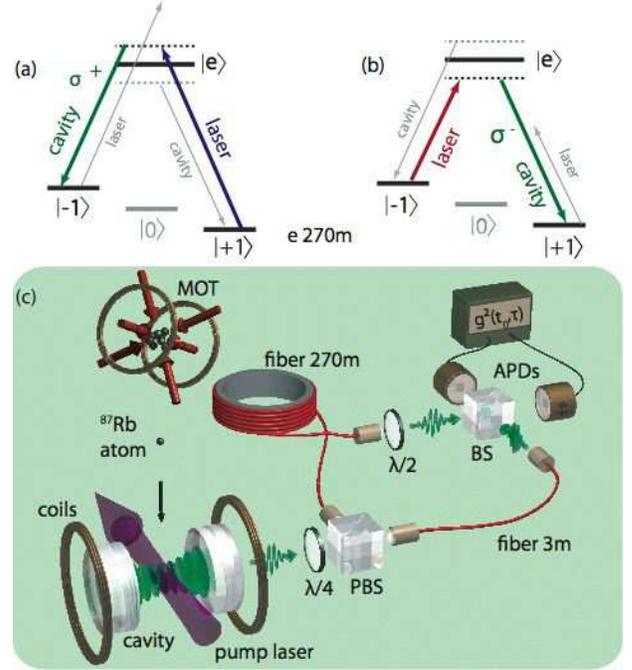}
 \caption{Scheme and setup: (a+b) Relevant energy levels in \rb{87} for photon production. (a) The pump laser and cavity drive a resonant Raman transition which takes the atom from \ket{+1} to \ket{-1}, producing a $\sigma^+$ photon. (b) A second pump pulse of different frequency returns the atom to \ket{+1}, whilst producing a $\sigma^-$ photon. (c) As atoms fall from a magneto-optical trap (MOT) through the cavity we shine in the pump laser from the side to generate photons.  The photons are directed through one of two fibers by a polarizing beam splitter (PBS), the long 270m fiber acting as a delay line. Photons emerging from the fibers can interfere at a beamsplitter (BS), and are detected by a pair of avalanche photodiodes (APDs).}
 \label{fig:setup}
\end{figure}

Figures \ref{fig:setup}(a) and (b) show the relevant levels involved in our scheme, the $5S_{1/2},\:F=1$ ground states, labelled by magnetic sublevel \ket{-1}, \ket{0}, and \ket{+1}, and the $5P_{3/2},\:F'=1,\:m_F=0$ excited state, \ket{e}. A magnetic field along the cavity axis defines the quantization axis and lifts the degeneracy of the ground states, producing an energy shift of the $m_F=\pm1$ states of $\mp\hbar\Delta_B$.  For geometrical reasons the cavity supports only $\sigma^+$ and $\sigma^-$ photons along its axis. The cavity resonance frequency $\omega_c$ is chosen to be $\omega_{0e}$, the frequency of the $\ket{0}\leftrightarrow\ket{e}$ transition, and is fixed during the experiment.   The pump laser is linearly polarized perpendicular to the quantization axis, decomposing into $\sigma^+$ and $\sigma^-$ components, and has a frequency $\omega_p$.  

Consider an atom, coupled to the cavity, in the \ket{+1} state. If the pump-cavity detuning $\Delta_{pc}\equiv\omega_p-\omega_c=2\Delta_B$ then the combination of a pump pulse with the cavity vacuum field resonantly drives a Raman transition and transfers the atom to the \ket{-1} state, depositing a $\sigma^+$ photon into the cavity.  If the cavity field decay rate $\kappa$ is similar to the Rabi frequency of the Raman transition, then the photon escapes from the cavity as the population transfer from \ket{+1} to \ket{-1} takes place.  Once the atom is in \ket{-1} and the photon has escaped, no transition to any other state is resonantly driven, ensuring that only a single photon is generated. 
To produce another photon with the next pump pulse, a pump-cavity detuning $\Delta_{pc}=-2\Delta_B$ is chosen to fulfil the Raman resonance for the transition from \ket{-1} to \ket{+1}.
 Then laser and cavity change their roles, resulting in the production of a single $\sigma^-$ photon. In contrast to previous experiments \cite{02:kuhn,04:keller,04:mckeever} this back and forth process requires no repumping laser pulse to return the atom to its initial state after a photon generation .

Figure \ref{fig:setup}c shows a schematic of the experiment.  \rb{87} atoms are dropped from a magneto-optical trap (MOT) through the TEM$_{00}$ mode of an optical cavity. The flux of atoms through the cavity mode is $\sim2/\ms$.  The cavity is 1\mm~long, has a finesse of 60,000, and the mode has a waist of 35\um.  One mirror has a 25 times larger transmission coefficient than the other.  The relevant parameters for the system are $(g_{max},\kappa,\gamma)/2\pi=(3.1,1.25, 3.0)\MHz$ where $g_{max}$ is the atom-cavity  coupling constant on the transition relevant for photon generation for an atom maximally coupled to the mode, and $\kappa$ and $\gamma$ are the field decay rates for the cavity and atom respectively.

A magnetic field of 20\G~along the cavity axis produces a Zeeman splitting of $\Delta_B/2\pi=14\MHz$. To generate a stream of photons, a sequence of pump laser pulses with alternating frequency is continuously repeated as the atoms fall through the cavity.  The Rabi frequency of the pump pulses goes as $\Omega_0{\rm sin}^2(\pi t/t_p)$, $\Omega_0$ being the peak Rabi frequency and $t_p$ the length of the pulse.  These pulses propagate perpendicular to both the motion of atoms and the cavity axis (and thus the magnetic field).

Photons produced within the cavity decay with a probability of 93\% through the mirror of higher transmittance. The photons have a well defined polarization so we can use a waveplate and polarizing beam splitter (PBS) to direct them into one of two polarization-maintaining single-mode optical fibers, one long (270\m), the other short (3\m).  The output modes of the fibers are recombined at a 50:50 non-polarizing beam splitter (BS), and photons are then detected at each output port by avalanche photodiodes. This detection setup allows us to characterize our source. Firstly, the single-photon nature is investigated when a fraction of photons is sent through only one fiber, while the other fiber is closed. In this case, we have a Hanbury Brown and Twiss setup allowing measurement of the intensity correlation as well as the detection-time distribution of the emitted photons. Secondly, the mutual coherence of pairs of $\sigma^+$ and $\sigma^-$ photons is characterized in a time-resolved two-photon interference experiment \cite{03:legero,04:legero}, where both fibers are used. The long fiber is used as a delay line for a first photon and the subsequently generated second photon is sent through the short fiber so the two photons arrive at the beamsplitter simultaneously, and can interfere.

\begin{figure}  
\centering
 \includegraphics{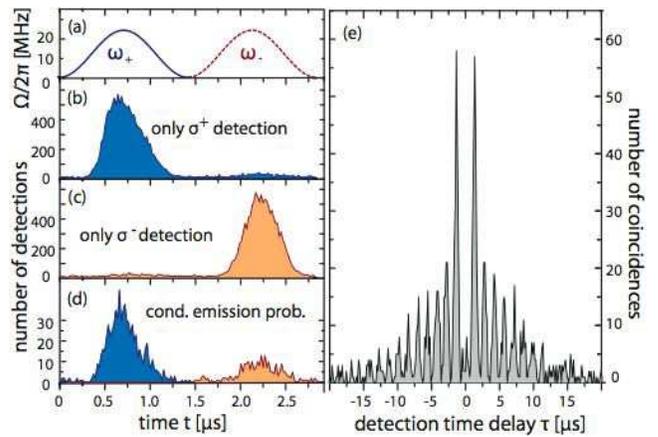}
 \caption{Photon characteristics observed with only one path to the beam splitter open. (a) Pump pulse sequence.  The first pulse, labelled $\omega_+$, has a detuning of $\Delta_{pc}/2\pi=28\MHz$, the second ($\omega_-$) $\Delta_{pc}/2\pi=-28\MHz$. (b)/(c) Photon detection time distribution for $\sigma^+$/$\sigma^-$ photons, showing they are overwhelmingly detected during the $\omega_+$/$\omega_-$ pulses. (d) Detection time distributions for photons conditioned on a detection during the previous pulse. The probability of detecting a $\sigma^+$ photon after a $\sigma^-$ photon is much larger than the probability of the opposite case. (e) Measurement of the intensity correlation between the two detectors (both polarizations are detected).  The missing peak at $\tau=0$ results from  the single photon nature of the source. Data are binned in 150\ns~intervals.}
 \label{fig:arrival}
\end{figure}

We first focus on characterizing individual photons.  In the measurements shown in figure \ref{fig:arrival} the long fiber is closed, and we only detect photons which pass through the short fiber. Figure \ref{fig:arrival}a shows the repeating sequence of pump pulses used, with parameters $(\Omega_0/2\pi, t_p)=(24\MHz, 1.42\us)$ and detunings $\Delta_{pc}=\pm2\Delta_B$.  The pulses are labelled $\omega_+$ and $\omega_-$ indicating that the laser frequency is such that a $\sigma^+$ or $\sigma^-$ photon, respectively, should be emitted.  To show that the frequency of the pump laser pulse determines the direction of the Raman process, and therefore the polarization of photons, we look at the detection time distributions of photons of the two circular polarizations separately.  In figure \ref{fig:arrival}b the waveplate is oriented such that only $\sigma^+$ photons are detected, in figure \ref{fig:arrival}c only $\sigma^-$.  It can be clearly seen that the number of $\sigma^+$ photons generated during the $\omega_+$ pulse is much larger ($\sim20$ times) than  during the $\omega_-$ pulse.  Similarly for $\sigma^-$ photons, the number of detected photons during the $\omega_+$ pulse is $\sim30$ times smaller than during the $\omega_-$ pulse. 
In both cases approximately the same number of atoms pass through the cavity and similar total numbers of $\sigma^+$ and $\sigma^-$ photons are detected.

A measure of the efficiency of our photon generation process can be obtained by considering the probability for a photon emission given that a photon was detected during the previous pump pulse.  This condition ensures that an atom is coupled to the cavity and that it is in the correct internal state to emit a photon in the subsequent pulse. For this measurement photons of both polarizations need to be detected, so the waveplate after the cavity is oriented such that the PBS acts as a 50:50 beamsplitter for each polarization.

Shown in figure \ref{fig:arrival}d are the detection time distributions of photons detected during the $\omega_+$ pump pulse, given that a photon was detected during the previous $\omega_-$ pulse ($t<1.42\us$), and the arrival time distributions of photons detected during the $\omega_-$ pulse, given that a photon was detected during the previous $\omega_+$ pulse ($t>1.42\us$).  Immediately obvious is the large difference in the number of these conditioned $\sigma^+$ and $\sigma^-$ photons.  Taking into account the dark count rate and the overall photon detection efficiency, we obtain two conditional probabilities for generating a photon inside the cavity:  $p(\sigma^+|\sigma^-)=41\%$ for generating a $\sigma^+$ photon after a $\sigma^-$ photon, and $p(\sigma^-|\sigma^+)=13\%$ for generating a $\sigma^-$ photon after a $\sigma^+$ photon.

In addition to the large difference between these conditional probabilities for the two polarizations, it can be seen in figures \ref{fig:arrival}b,c and d that the envelopes of the single-photon wavepackets depend on the polarization. The peak of the $\sigma^+$ envelope occurs earlier in the pump laser pulse than the peak of the $\sigma^-$ envelope. This asymmetry between the two polarizations occurs even though the scheme shown in figure \ref{fig:setup}a+b looks symmetric, however only the directly relevant levels are shown there.  Additional levels present in the atom (principally the $F'=0$ excited state) break this symmetry.  Nonetheless, theoretical studies that take the effect of additional levels into account \cite{06:wilk} are still insufficient to explain the measured differences in conditional probabilities.

To prove that only single photons are generated, an intensity correlation measurement is performed.  As in the measurement of conditional probabilities, we send photons of both polarizations through the short fiber. After the fiber the 50:50 beamsplitter randomly sends each photon to one of two avalanche photodiodes.  Figure \ref{fig:arrival}e shows the number of coincidences in the two detectors recorded as a function of the time delay $\tau$ between the detections. The comb-like structure reflects the periodicity of the driving pulses, where the width of the comb is a consequence of the limited interaction time a falling atom has with the cavity mode.  The feature we are most interested in is that the peak at time $\tau=0$ is missing, meaning that we have a single photon source. The probability of obtaining multiple photons is 2.5\% that of single photons. Note that the peaks at $\tau=\pm1.42\us$ are at least 2.5 times higher than all the others since a pair of subsequent photons is more likely to be obtained than three or more photons in a row.

Many QIP applications of a single-photon source require indistinguishable photons \cite{97:cirac,99:cabrillo,03:browne,03:duan,01:knill}.  Here, the envelopes of the $\sigma^+$ and $\sigma^-$ photons are seen to be similar in figure \ref{fig:arrival}b+c but this tells us nothing about the spectral properties.  We test for indistinguishability by performing a time-resolved two-photon interference experiment \cite{03:legero, 04:legero, 06:legero}.  Two photons that simultaneously enter different entrance ports of a 50:50 beam splitter will always leave through the same output port if they are indistinguishable.  The degree of indistinguishability is measured by comparing the number of coincidences obtained when the pair of photons have parallel polarization with the case when they are perpendicularly polarized and thus completely distinguishable. 

We superpose pairs of subsequently generated photons at a beamsplitter.  Both fibers are open, and since the photons have a well-defined polarization, they can be directed into either the long or the short fiber. A waveplate at the exit of the long fiber is used to set the relative polarization of the two photons.  To maximize the number of events where two photons reach the beamsplitter at the same time, $\sigma^-$ photons are delayed while $\sigma^+$ photons, which have the higher conditional emission probability, are sent directly to the beamsplitter. The time between the pump pulses is set such that there is maximum overlap of the pulse areas of the two interfering photons, this is what we denote a simultaneous arrival.  As in previous experiments \cite{04:legero} we count the number of coincidences as a function of the detection time difference, $\tau$.

\begin{figure}  
\centering
 \includegraphics{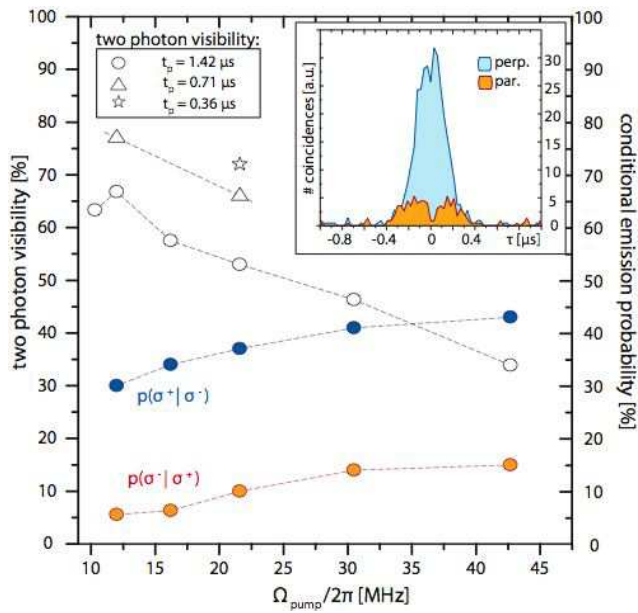}
 \caption{Two-photon interference.  Inset: coincidences as a function of the time difference between detections for photons with parallel and perpendicular polarizations.  For perfect spatial mode matching, the dip around $\tau=0$ for parallel polarizations would reach zero coincidences.  The observed dip is consistent with an interferometer visibility of 0.98.  $(\Omega_0/2\pi,t_p)=(12\MHz,0.72\us)$. An integrated visibility $V=77\%$ is observed. Main figure: visibility and photon-generation efficiency as a function of pump laser peak Rabi frequency $\Omega_0$ and various pulse durations $t_p$.  The open symbols refer to the integrated visibility $V$ of the two photon interference, with different shapes for different pulse durations.  $V$ increases with lower $\Omega_0$ and shorter $t_p$. The filled symbols show the conditional probabilities $p(\sigma^+|\sigma^-)$ and $p(\sigma^-|\sigma^+)$  for generating photons with a pulse duration $t_p=1.42\us$.}
 \label{fig:visibility}
\end{figure}

A typical result is shown in the inset of figure \ref{fig:visibility}.  The curves give the number of coincidences versus $\tau$ where the two photons have either parallel or perpendicular polarization at the beamsplitter. The dip in the number of coincidences for parallel polarizations around $\tau=0$ shows that the spatial mode matching of the two interfering photons is good, with the magnitude of the dip consistent with the measured single-photon interferometer visibility of 98\%. This means that most of the coincidences for parallel polarization reflect the distinguishability of the photons. 

We calculate the integrated two-photon interference visibility $V=1-(C_{\rm par}/C_{\rm perp})$ where $C_{\rm par}(C_{\rm perp})$ is the total number of coincidences for $|\tau|<t_p$ when the photons hit the beamsplitter with parallel(perpendicular) polarization. It is the same information one would get in a standard Hong-Ou-Mandel measurement \cite{87:hong} without detection time resolution at the bottom of the `HOM-dip'.  This visibility was measured for photons generated with a range of different peak Rabi frequencies $\Omega_0$, and pulse lengths $t_p$, the results of which are shown in figure \ref{fig:visibility}.  The visibility increases with reduced $\Omega_0$ and shorter pulse durations $t_p$, with a maximum measured visibility of 77\% for $\Omega_0/2\pi=12\MHz$ and $t_p=0.71\us$. The visibility is influenced by both the indistinguishability of the two photons and the spatial mode-matching at the beam splitter.  The combination of the interferometer visibility and slightly non-identical photon envelopes give a maximum possible visibility $V=94\%$, which would be obtained for interfering single-photon wavepackets with identical temporal evolution, higher than the maximum we obtained.

There are several processes that might affect the photon generation in a way that would make the photons more distinguishable. As discussed before, the atom is more complex than a three-level system, with additional levels in both the excited and ground states. Off-resonant transitions to these levels lead to frequency broadening with increasing Rabi frequency. In contrast to that, shorter photons show a better visibility as they have less time to dephase \cite{06:legero}.
As expected and shown in figure \ref{fig:visibility}, the conditional probabilities of generating a photon also change with $\Omega_0$ and $t_p$.  Unfortunately, conditions which lead to higher visibilities result in lower probabilities of generating photon pairs.  This reduces the rate at which the source could be used in QIP applications.  Of course, it is always possible to increase the visibility by post-selecting only those pairs of coincidences that occur within the dip around $\tau=0$.

In conclusion, we now have a way to generate single photons of known polarization in a well-defined spatial mode from a coupled atom-cavity system. The next step will be to use an atom trapped within the cavity \cite{05:nussman} to produce long streams of photons, with the same coupling to the cavity for each pulse.

\begin{acknowledgments}
We thank F. Verstraete for useful discussions.  This work was supported by the Deutsche Forschungsgemeinschaft (SFB 631, Research Unit 635) and the European Union [IST (QGATES, SCALA) and IHP (CONQUEST) programs].
\end{acknowledgments}

\end{document}